\documentclass[useAMS, usenatbib]{mn2e}
\usepackage{amsmath}
\usepackage[dvips]{graphicx}
\usepackage{graphicx, epsfig}
\usepackage{graphics}

\newcommand{\beq}{\begin{eqnarray}}
\newcommand{\eeq}{\end{eqnarray}}

\newcommand{\apj}{ApJ}

\newcommand{\mnras}{MNRAS}

\newcommand{\nat}{Nature}

\title[The Coarse Geometry of Merger Trees in $\Lambda$CDM]
      {The Coarse Geometry of Merger Trees in $\Lambda$CDM}
\author[J.E. Forero-Romero]
       {Jaime E. Forero-Romero $^{1,2}$\thanks{E-mail:jforero@aip.de}\\
$^{1}$Astrophysikalisches Institut Potsdam, an der Sternwarte 16, D-14482
Potsdam, Germany\\ 
$^{2}$Universit\'e Claude Bernard Lyon 1, CNRS UMR 5574, ENS Lyon, Centre de
Recherche Astrophysique de Lyon, \\Observatoire de Lyon,
9 Avenue Charles Andr\'e, 69561 St-Genis-Laval Cedex, France}
\begin{document}


\pagerange{\pageref{firstpage}--\pageref{lastpage}} 

\maketitle
\label{firstpage}

\begin{abstract}
We introduce the contour process to describe the geometrical properties of merger
trees. The contour process produces a one-dimensional object, the contour
walk, which is a translation of the merger tree. We
portray the contour walk through its length and action. The length is
proportional to to the number of progenitors in the tree, and the action can
be interpreted as a proxy of the mean length of a branch in a merger tree. 

We obtain the contour walk for merger trees extracted from the public database of the
Millennium Run and also for merger trees constructed with a public Monte-Carlo code
which implements a Markovian algorithm. The trees
correspond to halos of final masses between $10^{11} h^{-1}$ M$_\odot$ and
$10^{14} h^{-1}$ M$_\odot$. We study how the length and action of the walks
evolve with the mass of the final halo. In all the cases, except for the
action measured from Markovian trees, we find a transitional scale around
$3 \times 10^{12} h^{-1}$ M$_\odot$. As a general trend the
length and action measured from the Markovian trees show a large scatter in
comparison with the case of the Millennium Run trees.

\end{abstract}


\section{Introduction}
The current paradigm of large scale structure formation in the Universe is
hierarchical, meaning that structures are formed by merger aggregation.

The dominant driver for this merger dynamic is thought to be the non-baryonic
and non-collisional cold dark matter. Therefore, research in large scale structure
formation is principally done through the study of dark matter structures,
which on large scales are thought to serve as scaffolding for baryonic
structures \citep{LSS_SFW}. 

The dark matter merger process has been usually represented by a tree, keeping
the analogy with the genealogical tree of an individual. Great effort has been
put into the methods of construction of merger trees, as they are a way to
understand dark matter aggregation and are a necessary input for codes of
semi-analytic galaxy formation. The methods range from the analytical approach
of Monte-Carlo methods (\cite{2007IJMPD..16..763Z} for a review) passing through hybrid
approaches that mix numerical realizations of a density field with analytical
approximations for its evolution \citep{MORGANA} to the fully numerical
approach that identifies the dark matter halos from different snapshots in a
$N$-body simulation to construct the merger history\citep{galicsI}.  

Usually, the validity of the trees constructed in the analytical and hybrid
way is stated from comparisons with trees constructed from numerical
simulations \citep{1999MNRAS.305..946S,2007arXiv0708.1382P,2007arXiv0708.1599N}. Unfortunately, the
quantities used to compare trees from two approaches usually sacrifice the
complexity inherent to the tree structure in the sake of simple tests. The
most common simplification is to select one branch of the tree (the most massive)
to make the analysis. Another approach, measures the abundance of
structures of a given mass among the halos in all the merger tree. In all
these cases the geometrical information is suppressed, mostly because of the
lack of simple structures to describe that kind of information. 

In this paper we present a new way, in the astrophysical context, to translate
the geometrical information from a merger tree into a 1-dimensional
structure. The translation is based on the encoding of the tree information
into its \emph{contour walk}.  

We apply this description to the merger trees from a large dark
matter numerical simulation, the Millennium Run
\citep{Millennium}. We use its public database to select halos in different
bin masses at redshift $z=0$ to extract its merger histories and build the
contour walk. We analyze each tree in terms of two simple statistics extracted
from these walks.

We have also performed this kind of analysis on merger trees obtained with the
algorithm described by \cite{2007arXiv0708.1599N}, using the source
code they kindly made public. With this code we have constructed trees with
two resolutions for the minimum mass halo, one mimicking the Millennium Run,
and the other with nearly $8$ times lower resolution.

This paper is divided as follows. In Section 2 we explain the construction of
a contour walk from a merger tree. In Section 3 we present the simulation that
produced the public data we used in this paper and the Monte Carlo code for the
construction of Markovian merger trees. In Section 4 we perform an
immediate implementation of these concepts to the available merger trees.  
We calculate global statistics from the walks, and discuss its possible
physical meaning. In the last section we discuss
our results and suggest how the tool we have proposed can be used to tackle more
complex questions about merger trees.

\section{Trees and Walks}
\label{sec:trees}
The merger trees in the current dark matter paradigm are simple, as only
merging is allowed. From the mathematical point of view this tree structure
corresponds to a Galton-Watson tree. 

 Galton-Watson trees are genealogical trees of a population that
evolves according to very simple rules. At generation $0$, the
population starts with one individual, called the ancestor. Then each
individual in the population has, independently of the others, a
random number of children according to an offspring probability
distribution. In the structure formation context we identify the ancestor with
the halo at redshift zero, and the offsprings with the parents of the halo. 

These trees can be coded by a discrete
path around the contour of the tree \citep{legall}. This \emph{contour process} is easy to
visualize. We imagine the displacement of a particle that starts at time zero
from the root of the tree (a halo at redshift zero) and then visit the first
not yet visited parent halo, if any, and if none, the successor of the
halo. We do so until all the members in the tree have been explored and the
particle has come back to the origin. 

\begin{figure}
\includegraphics[scale=0.40, angle=180]{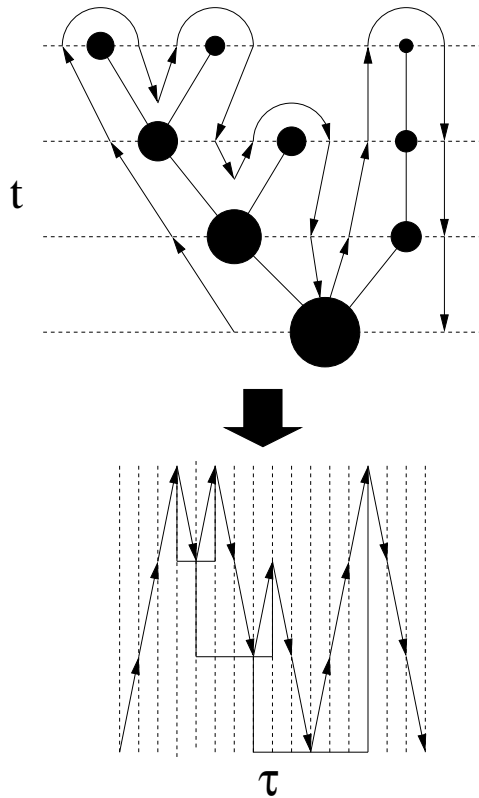}
\caption{Illustration of the contour process for a simple merger tree. The
  upper panel represents a merger tree formed by an evolving halo population at
  different timesteps $t$ (horizontal lines), time flowing from top to bottom.
  We start from the last halo at the bottom, and visit the
  halos in the tree at unit timesteps $\tau$. The most massive progenitor is
  always visited first. We go around until we go back to the initial point.  
  The bottom part of the figure shows the contour walk, its values correspond to
  the time $t$ where the particle was located at time $\tau $. The tree
  structure can be recovered completely from the contour walk as indicated by
  the solid lines in the lower panel.} 
\label{contorno}
\end{figure}

This process is illustrated in the Fig.\ref{contorno}.  Where the different
values of $t$ correspond to the discrete snapshots for which we have the halo
information, and $\tau_{i}$ is the imaginary time giving the pace of the
particle around the tree. We point out that one has to define some order
between the halos at a given point in the tree in order to define a unique way to walk
the tree, a criteria to decide to which parent should jump the particle. This
ordering criterium in our case is naturally imposed by the masses of the
halos, as we always visit first the most massive progenitor.

We can express this as visiting all the halos in the tree in a depth-first
fashion, visiting first the most massive branch at every time.

\section{N-body Simulation and Monte-Carlo code}
\subsection{The Millennium Run}
\begin{figure*}
\begin{center}
\includegraphics[scale=0.45]{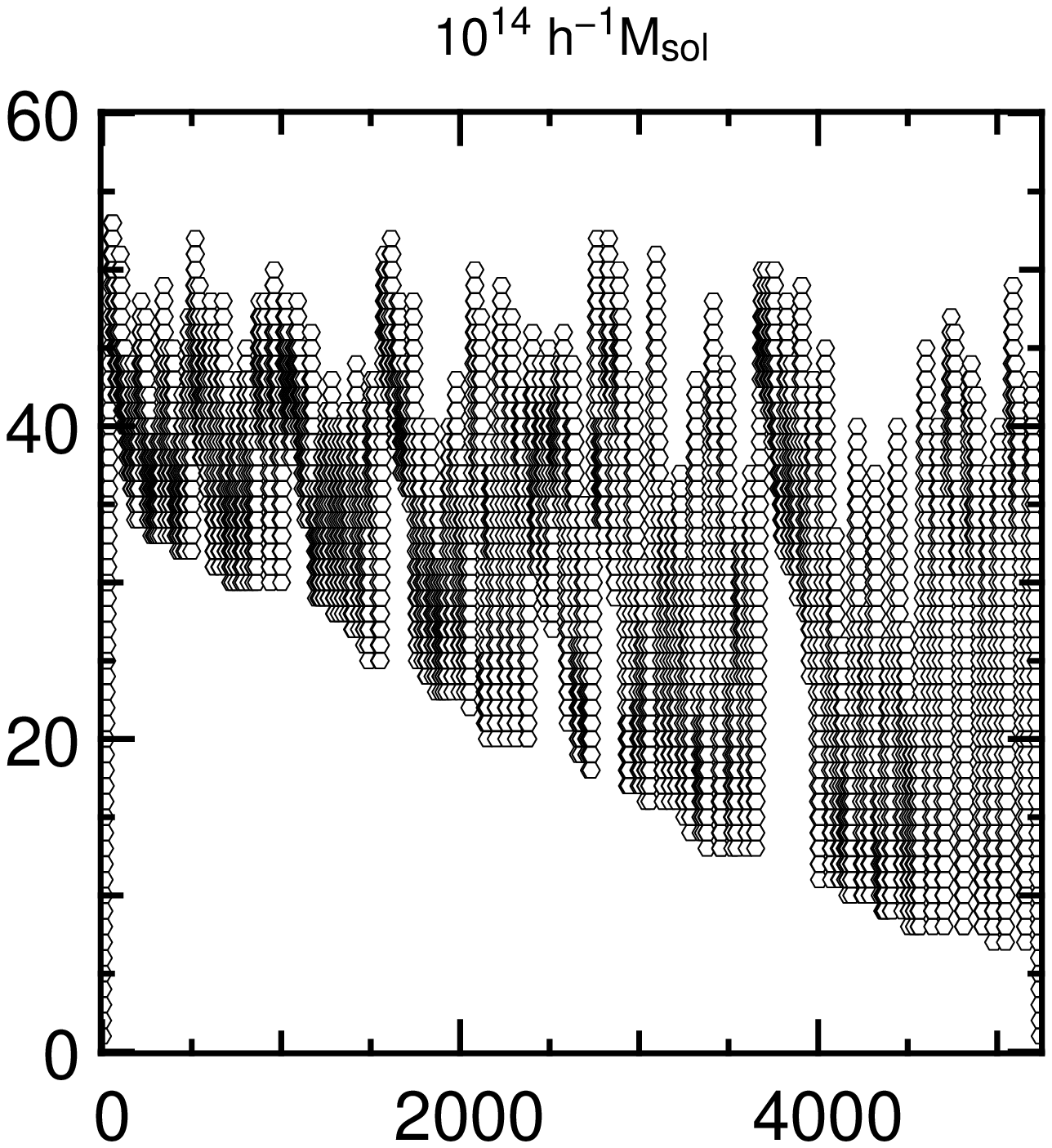}\hspace{0.5cm}
\includegraphics[scale=0.45]{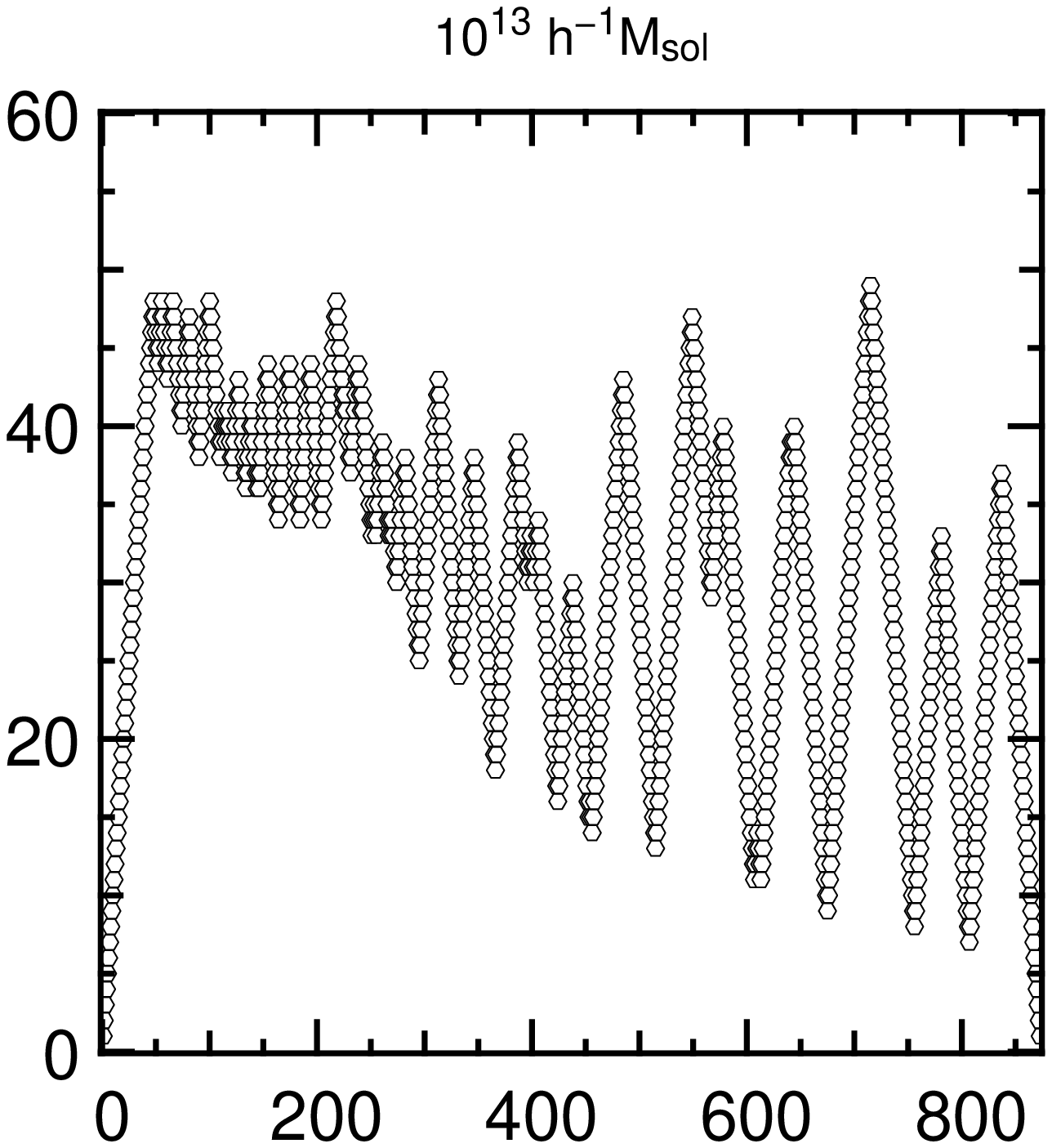}
\end{center}
\begin{center}
\includegraphics[scale=0.45]{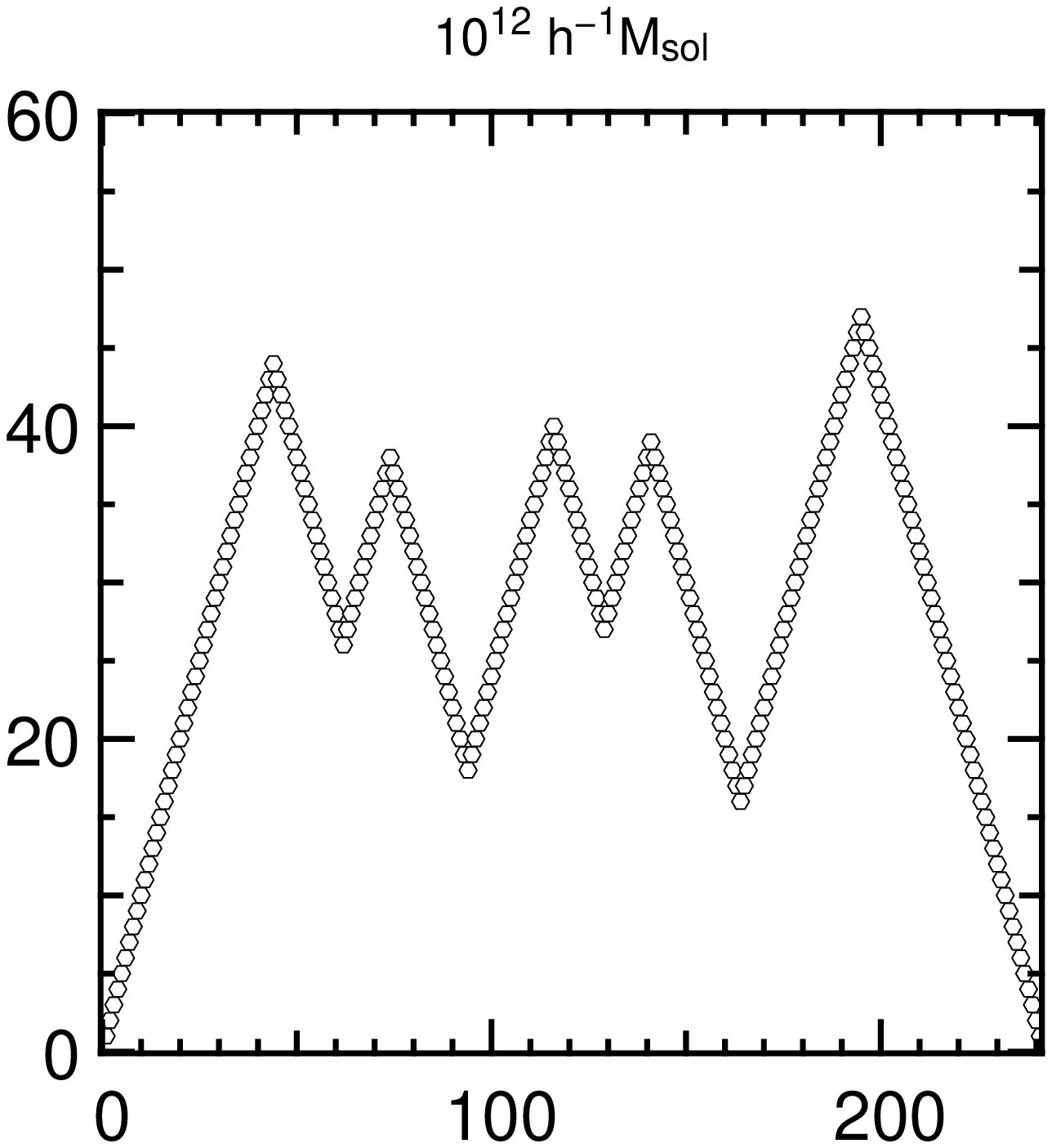}\hspace{0.5cm}
\includegraphics[scale=0.45]{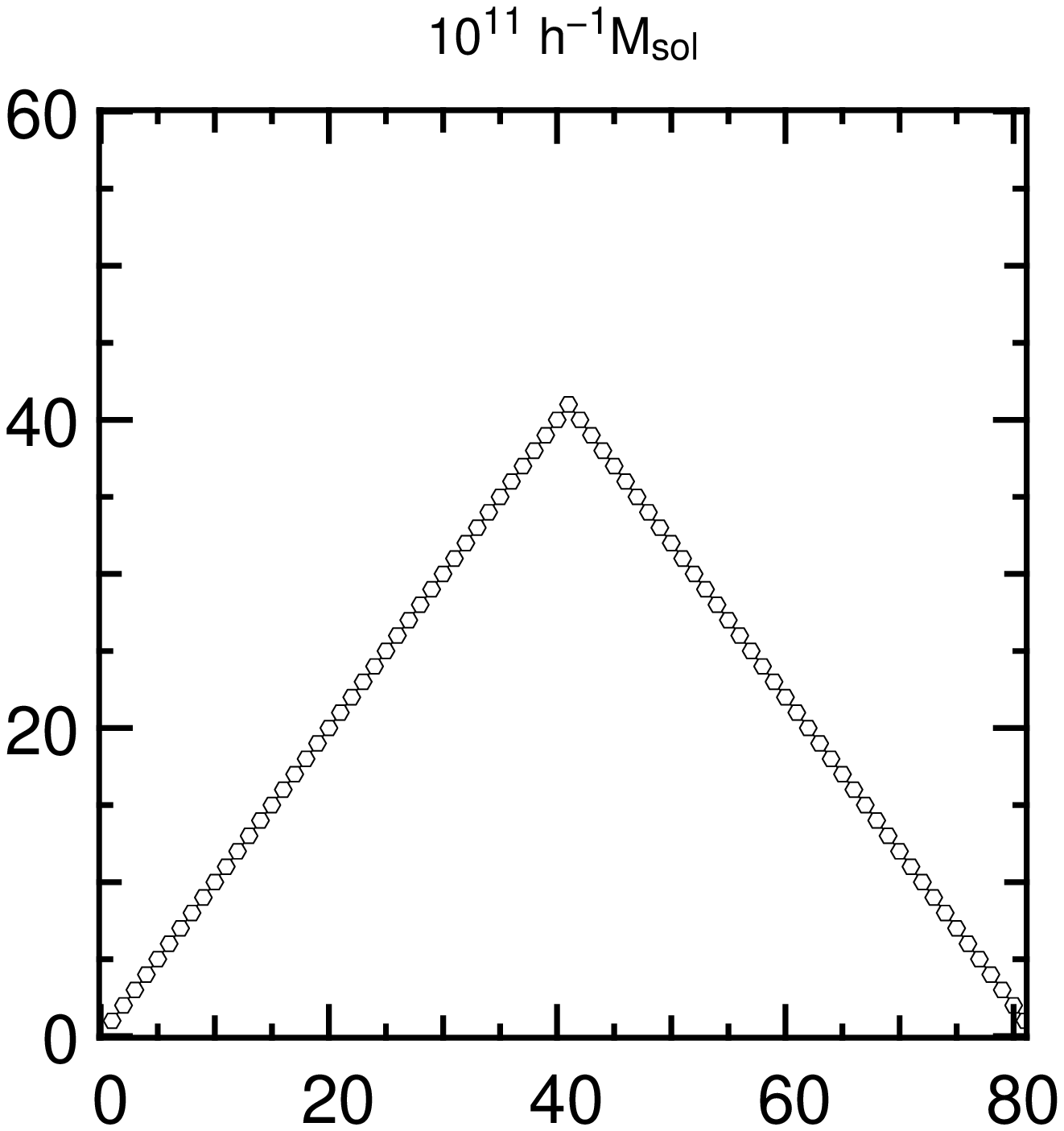}
\end{center}
\caption{\label{examples} Examples of contour walks constructed from merger
  trees extracted from the Millennium Run. Each contour corresponds to a halo
  of different mass. $\sim 10^{14} h^{-1}$ M$_{\odot}$ (upper left), $\sim
  10^{13} h^{-1}$ M$_{\odot}$ (upper right), $\sim 10^{12}h^{-1}$
  M$_{\odot}$ (lower left),  $\sim 10^{11}h^{-1}$ M$_{\odot}$ (lower
  right).}  
\end{figure*}

The Millennium Run \citep{Millennium} (MR hereafter) was carried out by the Virgo
Consortium in 2004 at the Max Planck's Society's supercomputer center in
Garching. 
It is a dark matter only simulation containing $2160^3$ particles, it was
evolved  from $z=127$ to the present-day inside a cubic region of $500$ Mpc
h$^{-1}$ on a side, the individual particle mass is $8.6\times 10^8 h^{-1}$
M$_{\odot}$. It adopted $\Omega_{dm}=0.205$,  $\Omega_{b}=0.045$ for the
current densities of dark matter and baryons in a Lambda Cold Dark Matter
cosmological model, furthermore 
it used $h=0.73$ for the present dimensionless value of the Hubble constant,
$\sigma_8=0.9$ for the \emph{rms} linear mass fluctuation in a sphere of
radius $8h^{-1}$Mpc extrapolated to $z=0$, and $n=1$ for the slope of the
primordial fluctuation spectrum.   

The simulation was run using the TREE-PM N-body code GADGET2 \citep{GADGET2}, 64
outputs were stored at times spaced approximately equally in the logarithm of
the expansion factor at early times and at approximately $200$ Myr intervals
after $z=1$. At run time all collapsed halos with at least $20$ particles
($1.7\times 10^{10} h^{-1}$ M$_{\odot}$) were identified using a friends-of-friends
(FOF) \citep{FOF} group-finder with linking length $b=0.2$. Post processing
with the substructure algorithm SUBFIND \citep{SUBFIND} allowed a detection
and measurement of the resolved sub-halos. This in turn allowed trees to be
built with detailed assembly histories for every object and its
substructure. 

The postprocessed data have been publicly disseminated through an interactive
database: {\tt http://www.g-vo.org/Millennium/MyDB}. The particular structure
of the database design which allows efficient querying for merger trees was
implemented by \cite{2006ASPC..351..212L}. Conveniently enough, the structure
for merger trees in the database is based on a depth-first ordering, making
that the output of a tree query is an incomplete version of the contour walk
In terms of the lower panel in Fig.\ref{contorno}, only the upward arrows
offspring-progenitor exist.

\subsection{Markovian Trees}
The authors \cite{2007arXiv0708.1599N} recently proposed and algorithm for the
construction of merger trees based on a Markovian approach. This 
approach to merger trees means that any 
halo of a given mass $M$ at time $t$ has a progenitor probability distribution
depending only on $M$ and $t$. This scheme is explicitly independent of the
large scale environment that could be defined for a halo.

Their merger trees are parameterized by a time variable $\omega\equiv
\delta_{c}/D(z)$ and a mass variable $S(M)=\sigma^2(M)$, where $\delta_{c}\sim 1.69$ and $D(z)$
is the cosmological linear growth rate and $\sigma^2(M)$ is the variance of
the initial density fluctuation field, linearly extrapolated to $z=0$ and
smoothed using a window function that corresponds to a mass $M$. These two
variables are also the natural variables in the Extended Press-Schechter
formalism \citep{2007IJMPD..16..763Z}.

Their approach is the following. First, from the MR data they find the conditional
probabilities for the masses of the main progenitors at a past time
$t^{\prime}$ as a function of halo mass $M$ at a present time $t$. With this
conditional probabilities they build the main progenitor history which by
construction reproduce to a good extent the MR data. Then, they extended this
approach to the construction of a full merger tree. The extension included some additional
heuristic rules based again on the premise of a fair match to the total progenitor
number density $dN/dM$ from the Millennium Run data. Nevertheless, the lack of
a true physical motivation cannot ensure the reconstruction of the correct full
joint distribution of progenitors.

Even if N-body merger trees are not Markovian, the proposed
algorithm manages to reproduce some tree properties, specially those related
with the main progenitor but also the total mass distribution in all the
progenitors. The detected inaccuracies come from the fact that the estimation
of the average mass for the second progenitor is not reliable.

In spite of that, we have decided to use these author's public available code given
its explicit effort to reproduce the MR data.

\section{Experiment Setup and Notation}

\subsection{Tree Selection}

We make use of two kind of trees. Trees extracted from the MR
public data base and trees constructed with the public code implementing the
Markovian approach. In both cases we concentrate on halos with
masses greater than $10^{11} h^{-1}$M$_{\odot}$.

For the MR trees we selected all the halos in the simulation box with a
given mass $M_{H}=\log_{10}(M_{200}/10^{10}h^{-1}$ M$_\odot )$ at redshift $z=0$, for $28$ different
values of $M_{H}$ in bins of width $\Delta M_H = 0.002$ dex, where $M_{200}$ is the halo mass
measured within the radius where the halo has an overdensity $200$ times the
critical density of the simulation. The bins are spaced by $0.1$ dex, the
least massive bin corresponds to $M_H=1.3$ and has $6028$ dark matter halos
and the most massive bin corresponds to $M_H=4.0$ and has $17$ dark matter
halos. We also performed measurements for the $15$ most massive bins
with $\Delta M_H=0.01$dex, which provided us with nearly five times more halos per
bin. The results we obtained in that case are basically the same than in the
case $\Delta M_H = 0.002$. In this paper we only report and discuss the
results for the selection with the smaller $\Delta M_H=0.002$ sample.

For the Markovian trees we made two different runs changing the minimal halo
mass. The first one mimicked the MR and its minimal mass was $1.7\times
10^{10} h^{-1}$ M$_{\odot}$. The second run used a higher minimal mass of $1.0\times
10^{11} h^{-1}$ M$_\odot$.  We will refer to these runs as the high resolution and
low resolution runs, respectively. In both cases we also constructed merger trees
for halo mass bins spaced by $0.1$ dex. The mass bins for the
high resolution run are the same as in the MR case, for the low resolution run
we can only describe trees from the bin $M_H = 1.6$ up to $M_H=4.0$.
For each bin we construct $1000$ trees, each one with $100$ steps in $\omega$ with
$\Delta\omega=0.1$. Which is equivalent to have trees described from redshifts
$z\sim 8$.

\subsection{Notation}

The snapshots in the dark matter simulation will be labeled by $t_{i}$, where
$i$ ranges from $0$ to $63$. The snapshot $t_{0}$ corresponds to the $z=0$. We
select a halo in the snapshot $t_{0}$ to extract its merger tree. The contour
walk, as described in Section \ref{sec:trees}, can be visualized as a
dynamical process of a particle going around the tree, at unit time-steps
stopping at each node in the tree and recording the snapshot $t_i$ to which it
belongs. The discrete variable counting the imaginary time of the particle
walking the tree is noted $\tau_i$.

We write the walk as a sequence of discrete values $\{x_{0},\ldots x_{N}\}$,
corresponding to different values of the discrete intermediary times
$\{\tau_{0}, \ldots , \tau_{N}\}$,  $N$ is the total length of the walk, and
every $x_{i}$ can take values from the possible snapshots $t_{0}\ldots
t_{63}$. In the case of the Markovian trees the values the walk can take vary
from $\omega_{0}$ to $\omega_{99}$, consistent with the fact that we have
described these trees at $100$ points equally spaced by $\Delta \omega=0.1$.

For each merger tree we compute two statistics: its length and
action. The length corresponds to the number of points in the walk. The
action, which will be defined later, is an statistic based on the first
derivatives of the walk.  We offer in Fig.\ref{examples},  a feeling on the
merger walks for halos in four different mass bins.

\section{Simple Statistics from the Walks}

\subsection{Walk Length}

\begin{figure}
\begin{center}
\includegraphics[scale=0.40]{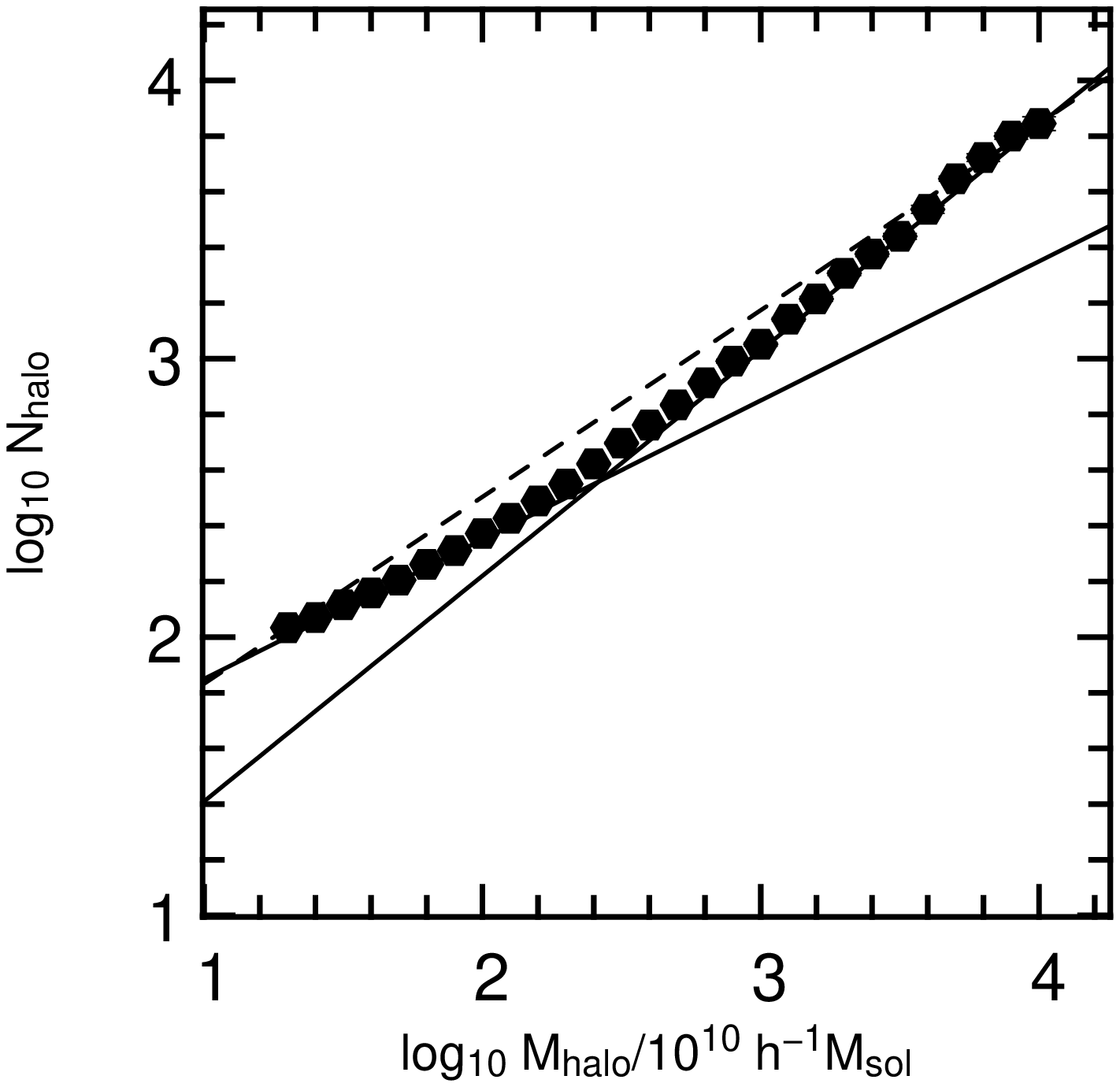}
\includegraphics[scale=0.40]{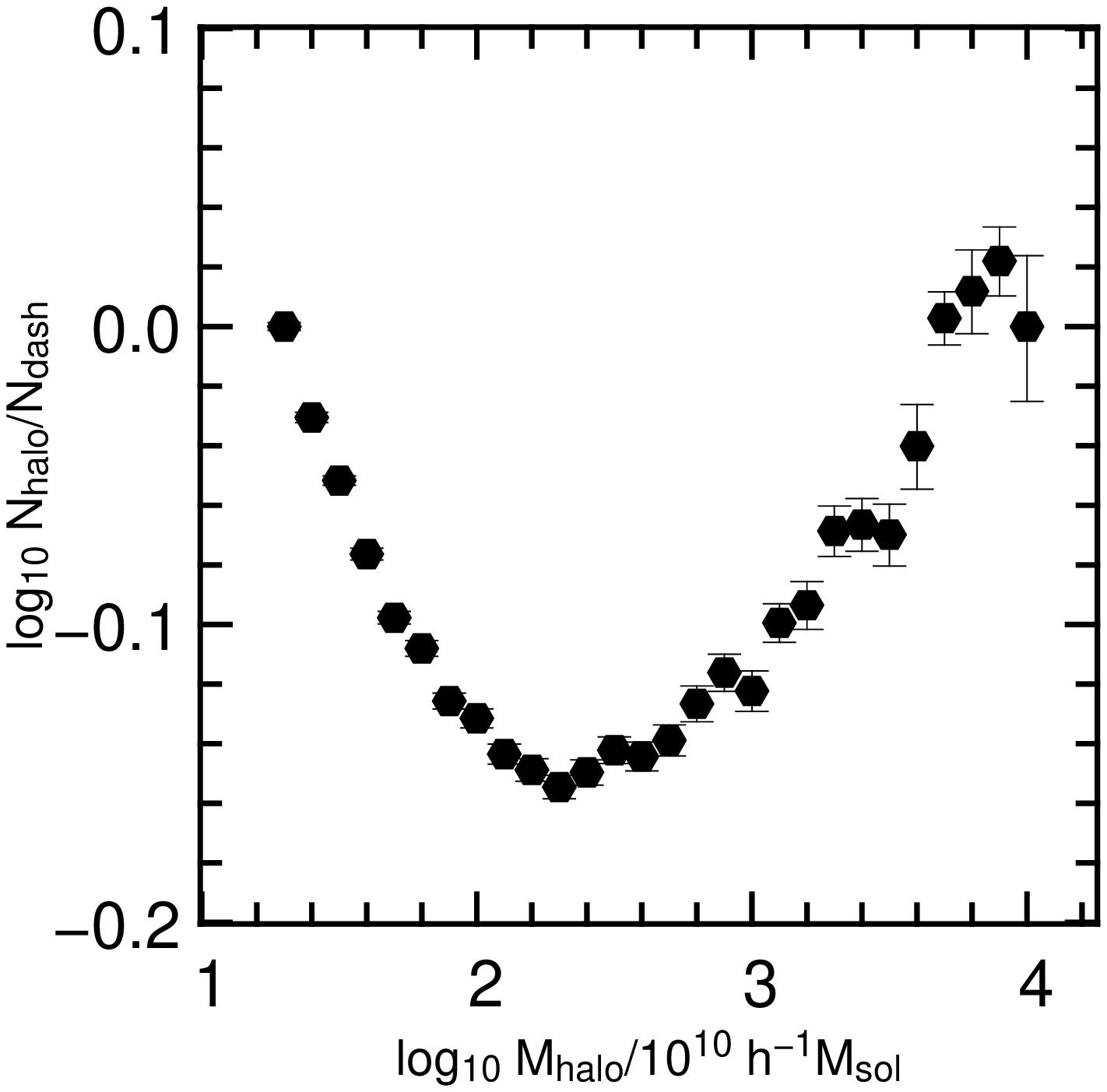}
\end{center}
\caption{\label{longhalo} \emph{Upper panel} Walk length $N_{halo}$ for halo merger trees extracted from  the Millennium Run as a function  of the logarithm of the final halo
  mass. The   solid lines show two different power law   trends for the
  relation $N_{halo}\propto   M_{halo}^{\lambda}$,   $\lambda=0.5$ and
  $\lambda=0.8$. The intersection of   the two lines is   located at $1.6\times
  10^{12}h^{-1}$M$_\odot$. The dashed   line simply   passes through the
  two extreme mass points in the plot.   \emph{Lower panel}   The same results
  as in the upper panel, but this time the   measured values   are normalized
  to the functional dependence of the dashed   line in the   upper panel. This
  enhances the features in the curve and allows   an easier   determination of
  the transitional mass-scale.}   
\end{figure}

The Figure \ref{longhalo} shows the walk length for halo merger trees as a
function of halo mass.  At low masses where the growth should be dominated by
mass accretion and not by mergers, $N\propto M^{0.5}$. At high masses where
the growth starts to be controlled by mergers $N\propto M^{0.8}$. The
transitional scale corresponds to halos for a halo mass $ 1.6\times10^{12}h^{-1}$M$_\odot$.

This transitional scale should depend on the resolution of the dark matter
simulation. When smaller halos are resolved, the merger trees will be
populated with more branches of this lower mass halos, and one could start to
see that all the growth is done trough mergers.  

Nevertheless, if we intend to study galaxy formation using merger trees, this
suggested transitional scale may have some significance. In the favored
paradigm of galaxy formation, not every halo can harbor galaxies at a cosmic
epoch. Only halos that can cool efficiently the gas may hold star forming
galaxies. A lower limit is imposed by the UV background from star formation,
which sets a low baryonic fraction in halos of masses below $\sim 1\times
10^{10}h^{-1}$M$_\odot$, which is roughly the minimal resolution for halo
detection in the Millennium Run. If one aims to study galaxy formation, the UV
background sets naturally a minimal mass of the progenitors that should be
included in the merger history of a dark matter halo. In any case, we decided
to explore the influence of the resolution using the results from the
Monte-Carlo code.  

\begin{figure}
\begin{center}
\includegraphics[scale=0.40]{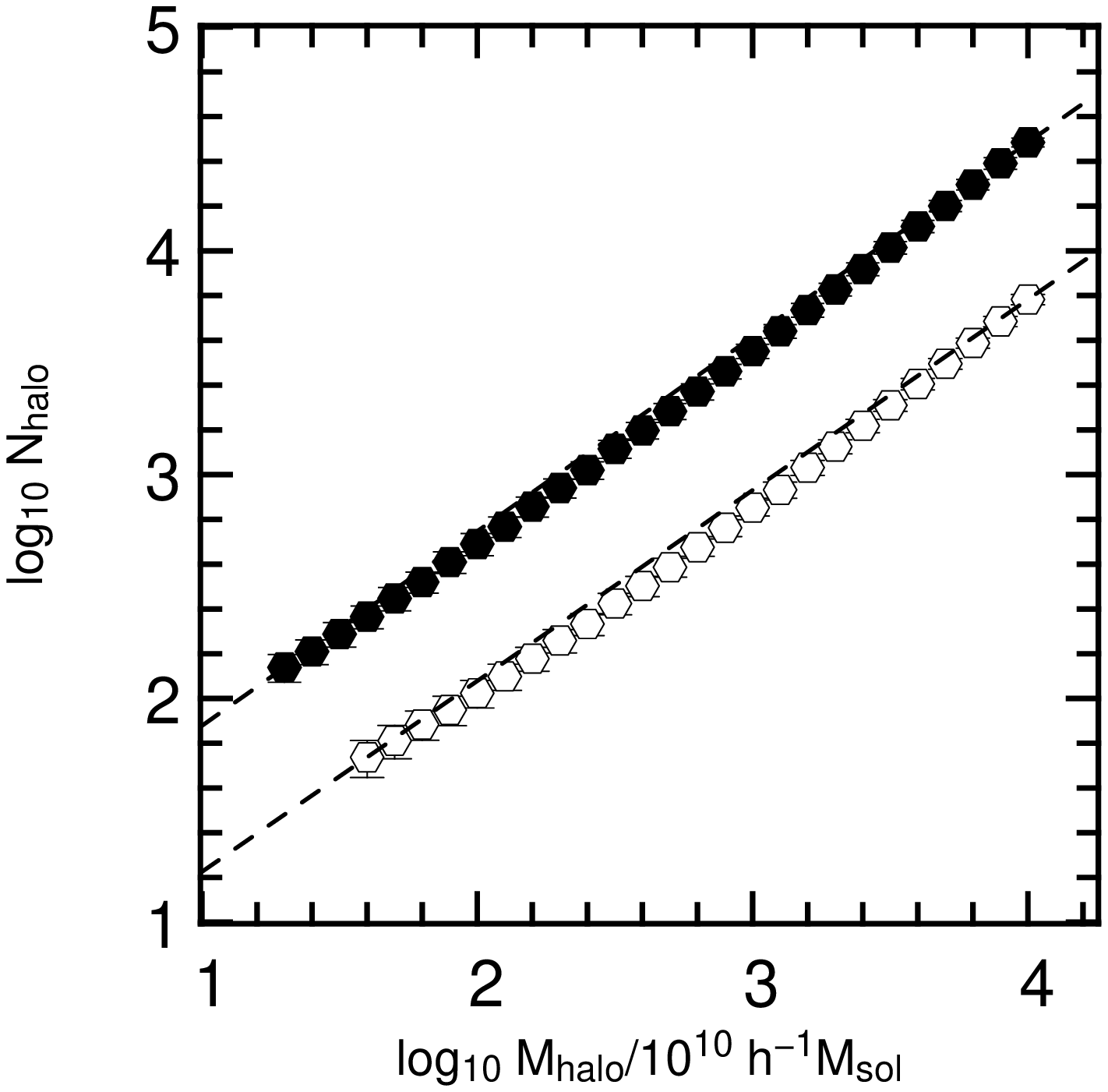}
\includegraphics[scale=0.40]{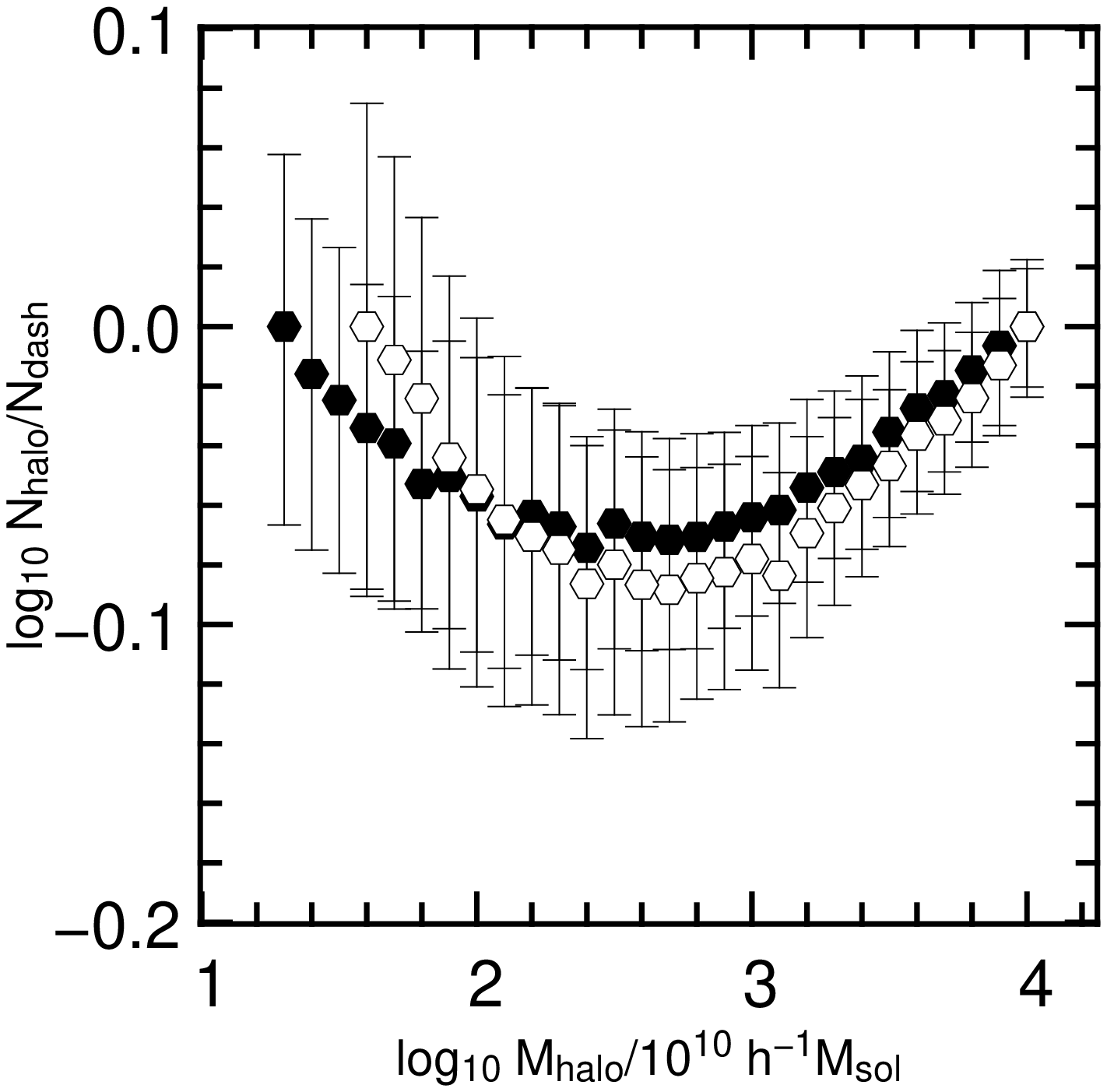}
\end{center}
\caption{\label{longhalo_markov} \emph{Upper panel}Same as Fig.\ref{longhalo} for merger trees
  constructed with the Markovian algorithm. Black symbols show the high
  resolution run, and white symbols the low resolution run. \emph{Upper panel}
  The
  lines show again two different power law trends for the relation $N_{halo}\propto
  M_{halo}^{\lambda}$, $\lambda=0.80$ and $\lambda=0.95$. This is more steeper
  than the $\lambda =\{0.5,0.8\}$ found in the Millennium merger trees. 
  \emph{Lower panel} The transitional mass-scale is roughly located at $4.0\times
  10^{12}h^{-1}$M$_\odot$ with a   weak dependence on the run's
  resolution. The dispersion for the measurements from the Markovian trees is
  noticeably larger than the measurements coming from Millennium Run data.} 
\end{figure}

In the case of merger trees constructed with  a resolution mimicking the MR we
find again a transitional scale (Fig.\ref{longhalo_markov}) around $4.0\times
10^{12}h^{-1}$M$_\odot$. The biggest difference in comparison with the
$N$-body results are the exponents describing the length of the walk as a
function of halo mass. In the Monte Carlo case both exponents are closer to
$\lambda\sim 0.9$. This results would favor a view from which there is still room left
for a better description of smooth accretion into the Markov description.

We extend the statistical description of the contour walk by the
quantification of the its waviness. We borrow from statistical mechanics
the concept of action, which is used in the numerical calculation of path
integrals \citep{BookKrauth}.

The action can be defined as the potential energy invested in bending the
path, comparing it with a characteristic temperature. If there is a walk
defined by points $\{x_{1}\ldots x_{N}\}$, the action is usually defined as
\begin{equation}
S = -\sum_{i=1}^{N}\frac{(x_{i+1}-x_{i})^2}{2\beta},
\end{equation}

where $\beta$ plays the role of a temperature or imaginary time depending on
the context. If we take that definition applied to our case:
\begin{equation}
S = \sum_{i=1}^{N} \frac{(x_{i+1} - x_{i})^2}{\tau_{i+1} - \tau_{i}}, 
\end{equation}

as we have unit steps, $\tau_{i+1} - \tau_{i}$ is equal to $|x_{i+1} -
x_{i}|$, making the action $S$ equal to the walk length. Therefore, we decide
to define a normalized action only from the extreme points in the contour walk. We take
this extreme points, $y_{i}$, as a sample of the contour walk where the
derivative at point $x_{i}$, defined as $(x_{i+1}-x_{i-1})/2$, is equal to
zero. We note the set of times $\tau_i$ corresponding to the vanishing points
of the derivative as $\mathcal{E}$ (as in $\mathcal{E}$xtreme). 

If we now define the action on the sampled walk $\{y_{i}|\ \tau_i\in \mathcal{E}\}$ 
\begin{equation}
S = \frac{1}{N_{E}}\sum_{i \in \mathcal{E}} \frac{(y_{i+1} - y_{i})^2}{\tau_{i+1} - \tau_{i}}
\label{accion}
\end{equation}

which makes the action $S$ equal to the walk length divided by the number of
extreme points. In the context of merger trees, each peak in the walk
corresponds to a branch in the tree, making the contribution to the action
roughly proportional to the length of that branch. Therefore, the normalized
action could be loosely interpreted as a proxy for the mean length of one branch in
the tree.

\subsection{Walk Action}
\begin{figure*}
\begin{center}
\includegraphics[scale=0.40]{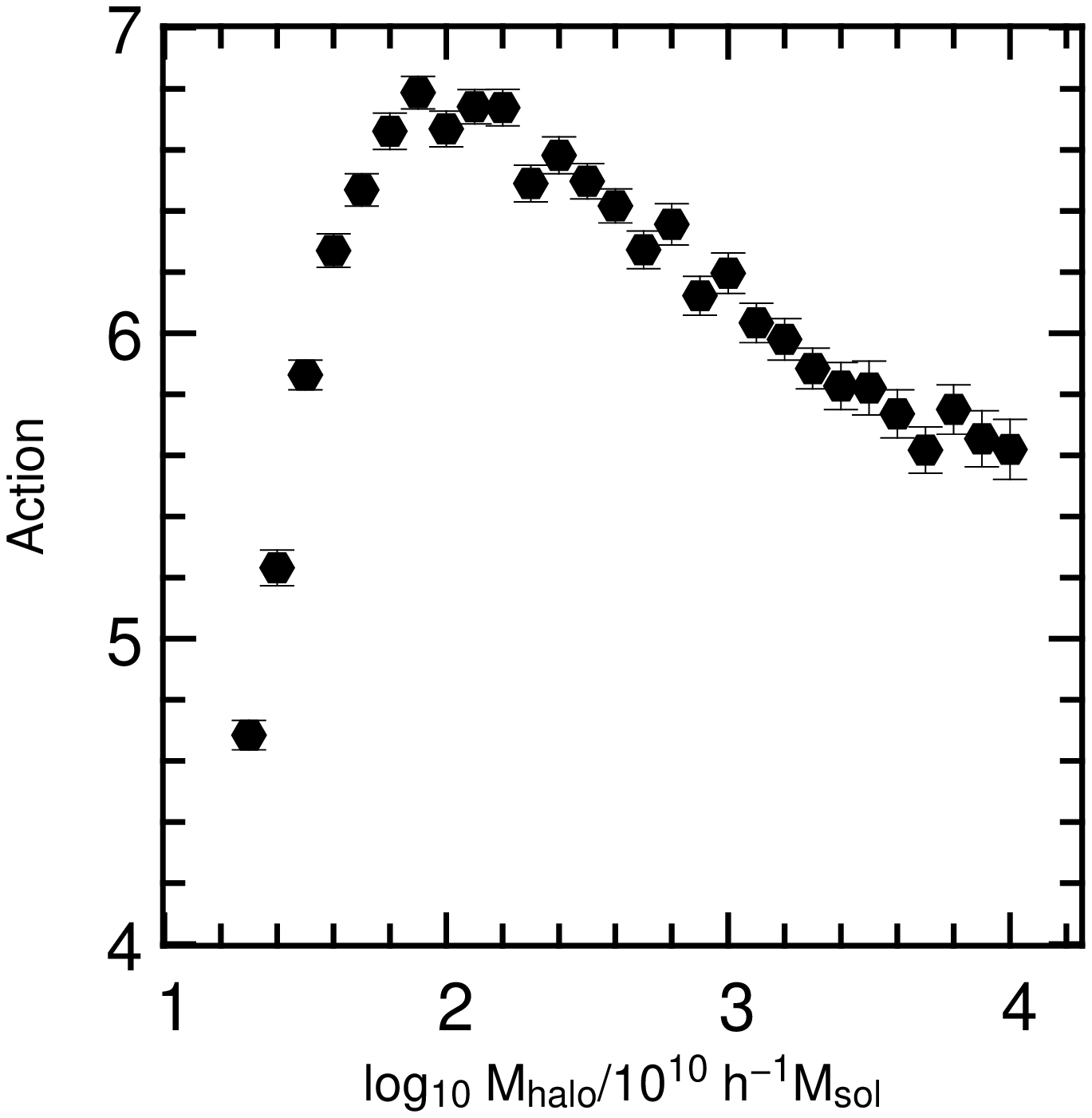}
\includegraphics[scale=0.40]{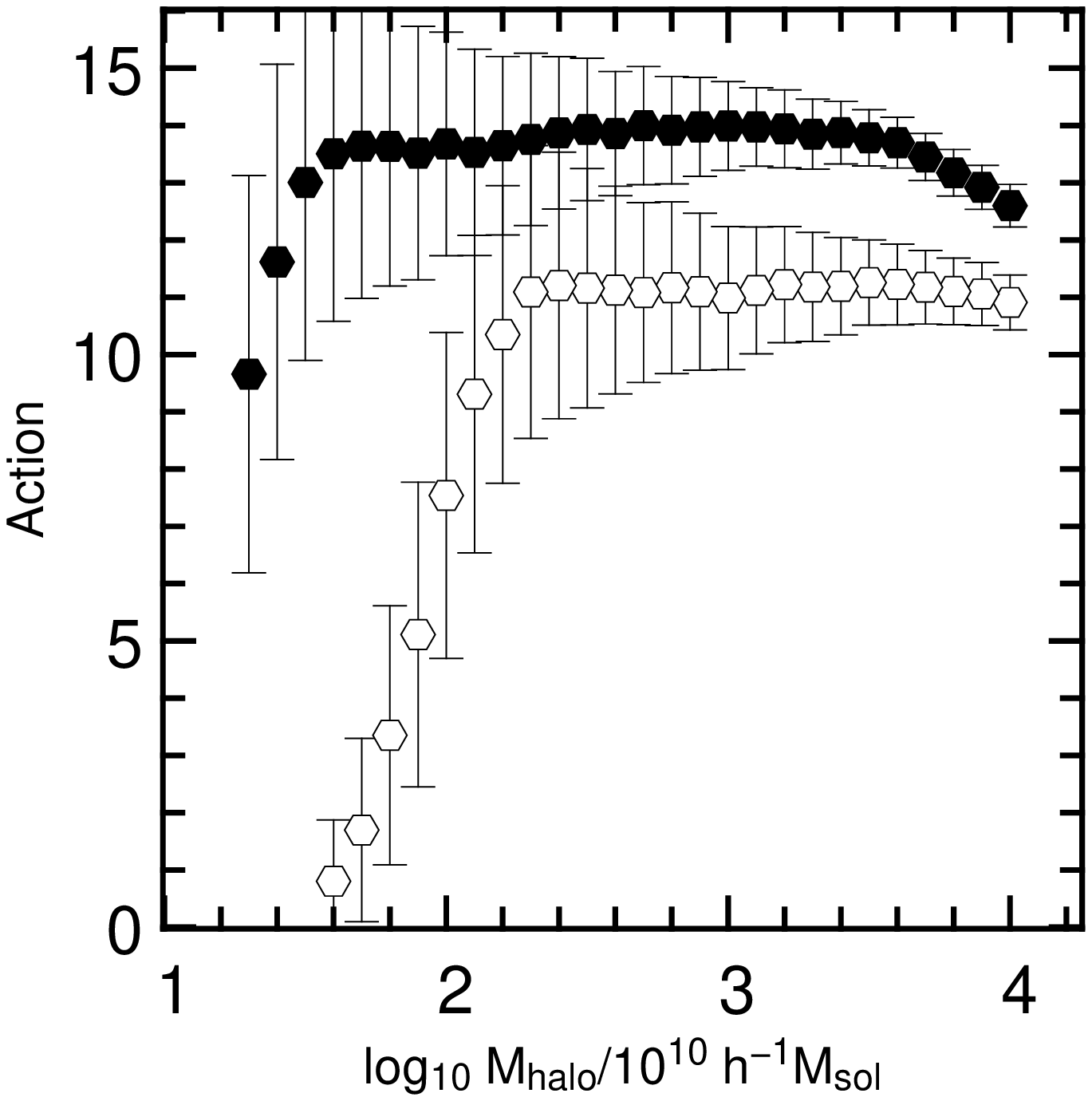}
\end{center}
\caption{\label{actionhalo} \emph{Left panel} Normalized walk action $S$,
  Eq.(\ref{accion}), as a function of halo mass for merger trees extracted
  from the Millennium Run. We find again a transitional mass-scale around
  $1.0\times 10^{12}h^{-1}$M$_\odot$, which could be interpreted as the mean branch
  length reaching a maximum for that halo mass. \emph{Right panel} The same as
  the left panel for merger trees constructed with the Markovian algorithm.
  Black symbols refer to the high resolution run, white symbols to the low
  resolution run. The transitional mass-scale is replaced by a plateau. Here
   the mass resolution plays an important role shaping
  the general trend for the action $S$. As in the length case, the dispersion
  in the measurements from Markovian trees is larger than the measurements
  coming from Millennium Run data.}
\end{figure*}

In Fig.\ref{actionhalo} we show the normalized
action as the function of halo mass for the MR and the Monte-Carlo code. For
MR trees we find a distinction between the behavior of the 
action for low and high mass halos. This time the two mass extremes share the
same action values, meaning that we can identify the same mean branch length
for the two extremes masses. For intermediate masses the action increases. In
the case of the MR trees, the action achieves  a maximum for a 
mass scale $1.0\times 10^{12} h^{-1}$ M$_{\odot}$. For  Monte Carlo trees the
maximum takes the form of a broader plateau ranging  almost two orders of
magnitude between $4.0\times 10^{11}h^{-1}$  M$_{\odot}$ and $4.0\times
10^{13} h^{-1}$  M$_{\odot}$.  

Perhaps the most distinguishing factor between the MR and Markovian approach is that
the dispersion in the latter is much higher that in the former. A fact that
can be interpreted as a higher variability in the geometry of Markovian  trees.

\section{Discussion}

We introduced from the mathematical literature the contour process of a tree,
 and we applied this concept to the
 description of merger trees. We used a large dark matter simulation (the
 Millennium Run) and a Monte-Carlo code (implementing a Markovian approach)
 to obtain merger trees  in these two approximations. Furthermore, the
 Markovian trees were obtained  with two different values for the minimal mass
 of a parent halo. One  resolution mimicked exactly the MR, and the other
 resolution had a $8$  times more massive minimal halo mass. We refer to these
 Markovian runs as the high resolution and low resolution runs respectively.

We extracted simple statistics from these walk: the length $N$ (proportional
to the total number of halos in the tree) and the action $S$ (which can be loosely
interpreted as the mean longitude of a branch). We report our results of walk
length and action emphasizing its evolution as a function of the physical halo
mass,  and not its absolute values.

From the length and the action, we found in the Millennium Run a transitional
mass scale at $\sim 3.0\times 10^{12} h^{-1}$ M$_{\odot}$. In the
case of the walk length, $N$, this transitional mass-scale marks the change
between a dependence $N\propto M^{0.5}$ for low halo masses and $N\propto
M^{0.8}$ for high mass halos, where $M$ is the mass of the final dark matter
halo in the tree. With the action, the scale marks the highest value for the
action as function of the halo mass.

For the Markovian trees the dependence of the walk length on the halo mass
is almost the same everywhere, although much steeper than the dependence
found in the Millennium Run. Nonetheless, we found the same transitional
scale from the length statistics for the two resolutions. It does not prove
that the transitional scale is independent on the resolution mass (in fact, it
should be dependent) but suggests that the scale does not have a strong
dependence with the minimal mass resolution used to describe the merger tree.
The evolution of the action as a function of halo mass is completely different
from the MR case. The Markovian run with mimicking MR resolution does not show a
sharp transitional scale, instead it shows a large plateau ranging for two
orders of magnitude in mass.  

Perhaps the biggest difference between the two ways of constructing the trees
is that the Markovian approach always show a bigger dispersion on its contour walk
statistics, which seem best defined in the Millennium Run, judging from its low
dispersion. This could be interpreted in fact as a higher geometrical
variability in the Markovian trees.

We have shown how simple statistics from the contour walk can give a new
handle on the description of merger tree, exploring the
geometrical information of merger trees. Even if the length of the walk could
have been obtained without an intermediating contour process, the approach
of extracting information from higher order statistics, as was the case for
the normalized action $S$, can be extended in complexity. For instance, is relevant to point out that the information of the mass in each
node is also encoded in the contour walk. The mass information is used as an ordering
criteria to walk the tree. One could define sections defined by
$\{\tau_{i}|x(\tau_{i}=t_{1})\}$, i.e. the points where the walk touches the
next to last snapshot, and the \emph{length ratio} of the sections delimited by these
points would include information of the \emph{mass ratio} of the mergers at time
$t_{1}$. In general the study of crossing paths, defined as the the section of
the contour walks within some boundary $a<\{x_{i}\}<b$ might produce useful
statistics to further classify the complex merger trees of massive halos.

\section*{Acknowledgments}
We thank J\'er\'emy Blaizot (JB) and Gerard Lemson (GL) for early discussions,
motivation and criticism around the ideas presented here. We thank JB's
suggestion on pointing the discussion towards a comparison with Markovian
merger trees. We thank again GL for his amazing day-to-day work on the public
Millennium database. We also thank the authors Neistein and Dekel for making
public a well documented and easy to use code for merger tree construction. The
necessary infrastructure to develop this work was provided in the framework of
the HORIZON Project (France).

\end{document}